\documentclass{svproc}
\usepackage{graphicx}%
\usepackage{multirow}%
\usepackage{amsmath,amssymb,amsfonts}%
\usepackage{mathrsfs}%
\usepackage[title]{appendix}%
\usepackage{textcomp}%
\usepackage{manyfoot}%
\usepackage{booktabs}%
\usepackage{algorithm}%
\usepackage{algorithmicx}%
\usepackage{algpseudocode}%
\usepackage{listings}%
\usepackage{url}

\usepackage{datetime}
\usepackage[dvipsnames]{xcolor}

\begin{document}
\mainmatter              
\title{Signature-Based Community Detection for Time Series}
\titlerunning{Signature-Based Community Detection for Time Series}  

\author{Marco Gregnanin \inst{1, 2} \and
Johannes De Smedt\inst{2}, Giorgio Gnecco \inst{1} \and
Maurizio Parton \inst{3}}
\authorrunning{M. Gregnanin et al.}

\institute{IMT School for Advanced Studies Lucca, Italy \and
KU Leuven, Belgium\\
 \and
University of Chieti-Pescara, Italy \\
}

\maketitle  
\begin{abstract}
Community detection for time series without prior knowledge poses an open challenge within complex networks theory. Traditional approaches begin by assessing time series correlations and maximizing modularity under diverse null models. These methods suffer from assuming temporal stationarity and are influenced by the granularity of observation intervals.\\
In this study, we propose an approach based on the signature matrix, a concept from path theory for studying stochastic processes. By employing a signature-derived similarity measure, our method overcomes drawbacks of traditional correlation-based techniques.\\
Through a series of numerical experiments, we demonstrate that our method consistently yields higher modularity compared to baseline models, when tested on the Standard and Poor's 500 dataset. Moreover, our approach showcases enhanced stability in modularity when the length of the underlying time series is manipulated.\\
This research contributes to the field of community detection by introducing a signature-based similarity measure, offering an alternative to conventional correlation matrices. 
\end{abstract}

\section{Introduction}
In recent years, the exploration of structural properties within complex systems has garnered considerable significance across diverse scientific domains, including biological, social, communication,  economical, and financial networks. Of particular interest is the investigation and identification of communities within such networks. In fact, identifying communities within a network provides information on how the nodes are connected and organized \cite{barabasi2013network}. Especially challenging is the task of community detection within financial time series data, given their temporal dependencies, inherent noise, and non-stationarity \cite{tsay2005financial}. Identifying stock communities is particularly important for portfolio strategies and risk management tasks \cite{prigent2007}.
Conventional methodologies for community detection involve transforming time series into a graph-based representation by filtering the correlation matrix \cite{Mantegna1999econophysics,sinha2010econophysics}. Common techniques encompass the utilization of a threshold, referred to as the ``Asset Graph'' approach \cite{onnela2004clustering,heimo2009mst}, or employing alternative embedding methods to reduce dimensionality. Notably, Random Matrix Theory (RMT) \cite{mehta2004rmt,bai2010rmt} is a prevalent approach, which filters the correlation matrix by identifying and eliminating eigenvalues associated with noise. However, reliance on the correlation matrix for time series representation introduces biases, which can be summarized, among other constraints, by assuming temporal stationarity, and focusing on linear relationships  \cite{brockwell2002timeseries}.\\
In this study, we propose a paradigm shift by substituting the correlation matrix with a similarity matrix derived from time series signatures \cite{lyons1998,lyons2014}. The rationale behind employing the signature, as opposed to the original time series, lies in its remarkable capability to encapsulate temporal information of the underlying time series within a systematically structured sequence of tensors \cite{chen1958,lyons2014path}. \\ 
The rest of the paper is organized as follows: Section \ref{section_2} provides an overview of core filtering techniques for correlation matrices and delves into the intricacies of community detection in the context of financial time series. Section \ref{section_3} summarizes the ``Asset Graph'' approach, RMT, community detection, and signatures. Section \ref{section_signature} defines the similarity matrix derived from the signature, Section \ref{section_4} provides a numerical experimental comparison, and Section \ref{section_5} concludes the paper. 

\section{Related Work}
\label{section_2}
Analyzing the structural properties and filtering techniques in correlation matrices of time series data can be approached through various methods. One straightforward approach is based on the ``Asset Graph'', wherein the correlation matrix is filtered using a threshold-based method \cite{onnela2004clustering,heimo2009mst}. Specifically, matrix elements are retained if they are greater than or equal to a 
given threshold value. However, the challenge lies in determining the optimal threshold value. Potential solutions include considering statistically significant correlation coefficients \cite{fisher1915} or setting the threshold by imposing that nodes within the same community have larger probabilities to be connected by edges than background edge probabilities \cite{yuan2022}. \\
Another filtering technique involves utilizing a Minimum Spanning Tree (MST) \cite{mantegna1999mst}. This method selects a subset of edges forming a tree that connects all nodes through the  links associated with the largest correlation. MST has been applied to filter the correlation matrix of the top $100$ capitalized stocks on the U.S. markets \cite{bonanno2004network}. Building upon the MST concept, the Planar Maximally Filtered Graph (PMFG) was introduced \cite{tumminello2005filtering}, which differs by retaining $3(n-2)$ links compared to MST's $n-1$ links, where $n$ represents the number of nodes. PMFG also allows for cliques and loops. PMFG was utilized to study the New York Stock Exchange's (NYSE) correlation matrix's topological properties \cite{tumminello2007correlation}.\\ 
The Random Matrix Theory (RMT) \cite{mehta2004rmt,bai2010rmt} offers another approach for filtering correlation matrices to extract meaningful information while removing noise. RMT has been employed to study eigenvalues' density and structural properties of empirical correlation matrices, such as those of Standard and Poor's $500$ (S\&P$500$) \cite{laloux1999noise} and the Tokyo Stock Exchange \cite{utsugi2004rmt}. Additionally, RMT has been applied to filter the Financial Times Stock Exchange (FTSE) index's correlation matrix for portfolio creation \cite{potters2005rmt}, to analyze eigenvalue properties and cluster stocks in the FTSE index and S\&P$500$ \cite{livan2011spectral}, and for community detection in the S\&P$500$ \cite{macmahon2015community}.\\
Other clustering techniques include the Potts method \cite{heimo2008modules}, which detects modules based on a dense weighted network representation of stock price correlations, and node-based clustering \cite{fenn2012clustering}, applicable to foreign exchange data and capable of tracking temporal dynamics. The Generalized Autoregressive Conditional Heteroskedastic (GARCH) model has been used to denoise Japanese stock return correlation matrices, followed by spectral clustering \cite{isogai2014clustering}. Clustering based on similarity in distribution of exchange rates in the international Forex market has been explored \cite{chakraborty2020}.\\
Finally, the community detection problem is typically addressed using two established algorithms: the Louvain Community Detection Algorithm \cite{blondel2008} and the Clauset-Newman-Moore Greedy Modularity Maximization Algorithm \cite{clauset2004}.

\section{Preliminaries}
\label{section_3}
In this section, we provide an overview of the primary methodologies employed for filtering correlation matrices.  Furthermore, we define the community detection problem and present the key algorithm utilized to address it. We subsequently introduce the concept of a time series signature and expand upon its derivation from time series data. \\
Consider a collection of $N$ univariate time series denoted as $S$, each consisting of realizations over $T$ discrete time steps, represented as $S_i = \{s_i(1), s_i(2), \dots, s_i(T)\}$.
The entries of the correlation matrix $C$ among the $N$ time series can be defined
as follows:
\begin{equation}
C_{ij} = \frac{\sigma_{S_i,S_j}}{\sqrt{\sigma^{2}_{S_i}}\sqrt{\sigma^{2}_{S_j}}}\,. \label{eq:correlation matrix}
\end{equation}
Here, $\sigma_{S_i,S_j}$ represents the covariance between time series $i$ and $j$, while $\sigma^{2}_{S_i}$ denotes the variance of time series $i$. These 
are expressed empirically as:
\begin{eqnarray*}
\sigma_{S_i,S_j} &=& \frac{1}{T} \sum_{t=1}^{T}s_i(t)s_j(t) - \frac{1}{T} \sum_{t=1}^{T}s_i(t)\frac{1}{T} \sum_{t=1}^{T}s_j(t)\,,\\
\sigma^{2}_{S_i} &=& \frac{1}{T} \sum_{t=1}^{T}s^{2}_i(t) - \frac{1}{T} \sum_{t=1}^{T}s_i(t)\,.
\end{eqnarray*}
The correlation matrix in this study will be constructed based on the logarithmic increments of the time series $S_i$. Logarithmic increments are defined as $r_{i}(t)=\log \left(\frac{S_i(t)}{S_i(t-1)} \right)$ for all $t$ belonging to the set $\{2,3,\ldots,T\}$.

\subsection{Asset Graph}
The Asset Graph is based on the utilization of a threshold-based approach for filtering the correlation matrix. Specifically, we retain those entries within the correlation matrix that are greater than or equal to a predefined threshold. The selection of an appropriate threshold can be approached through various methods. An initial method involves evaluating multiple threshold values and selecting the one that generates a sparsely filtered matrix without introducing excessive disconnected components. An alternative strategy, outlined in \cite{fisher1915}, centers on retaining only those correlation entries that possess statistical significance. 

\subsection{Random Matrix Theory}
The objective of this approach is to extract information from the correlation matrix by discerning and retaining the relevant components while excluding the noisy elements based on the eigenvalues' distribution. Consider a correlation matrix derived from a set of $N$ completely random time series, each with a length of $T$. Following the principles of RMT, when $N\rightarrow +\infty$, $T\rightarrow +\infty$, and $1 < \lim T/N < +\infty$, the eigenvalues of the correlation matrix follow the Marcenko-Pastur distribution \cite{laloux1999noise,plerou1999}, denoted with $\rho(\lambda)$:
\begin{equation}
    \rho(\lambda) = \frac{Q}{2 \pi\sigma^2}\frac{\sqrt{(\lambda_{+}-\lambda)(\lambda-\lambda_{-})}}{\lambda}\,,\ \ \ \ \ {\rm if}\ \ \lambda \in [\lambda_{-},\lambda_{+}]\,, \label{eq:rmt_distribution} 
\end{equation}
and zero otherwise. Here, $Q=\lim \frac{T}{N}$, $\lambda_{\pm}=\sigma^{2}\left(1\pm \sqrt{\frac{1}{Q}}\right)^2$, and $\sigma^{2}$ is the variance of the elements, often set empirically as $\sigma^{2} = 1 - \frac{\lambda_{max}}{N}$, where $\lambda_{max}$ represents the maximum eigenvalue of the correlation matrix. In RMT, eigenvalues greater than $\lambda_{+}$ are statistically significant, while the rest are largely attributable to random noise. As such, any correlation matrix can be decomposed into the sum of a structural component $C^{(s)}$ comprising eigenvalues above $\lambda_+$, and a noise component $C^{(r)}$ which can be expressed as:
\begin{equation}
C^{(r)}=\sum_{i:\lambda_i \leq \lambda_{+}}\lambda_iv_iv_i^{\dagger}\,. \label{eq:noise_component}
\end{equation}
Here, $v_i$ represents the eigenvector associated with eigenvalue $\lambda_i$, and $v_i^\dagger$ is its 
conjugate transpose. \\ 
However, in case of the empirical log-returns correlation matrix for $N$ stocks, an eigenvalue often is significantly greater than the rest, and is commonly referred to as the ``market mode'' \cite{sinha2010econophysics,mehta2004rmt,laloux1999noise}. The market mode encapsulates the market's overall behavior, impacting all other stocks. Consequently, removing the market mode is essential for enhancing the detection of valuable correlations by reducing noise interference. The correlation matrix for $N$ stocks can thus be decomposed into three components:
\begin{equation}
    C = C^{(r)} + C^{(m)} + C^{(g)}\,, \label{eq:correlation_decomposition1}
\end{equation}
where $C^{(r)}$ represents the noise component, $C^{(m)}$ is the market component, and $C^{(g)}$ is the remaining significant correlation, after the removal of noise and market mode components. Specifically:
\begin{eqnarray}
    C^{(m)}&=&\lambda_{max}v_{max}v_{max}^{\dagger}\,, \label{eq:market_component}\\
    C^{(g)}&=&\sum_{i:\lambda_+ < \lambda_i < \lambda_{max}}\lambda_iv_iv_i^{\dagger}\,, \label{eq:real_component}
\end{eqnarray}
where $v_{max}$ represents the eigenvector associated with the maximum eigenvalue of the correlation matrix. 
Finally, the filtered correlation matrix utilized for the community detection problem (see the next subsection) is $C^{(g)}$. 

\subsection{Community Detection}
Community detection aims to identify groups of nodes within a network that are more likely to be interconnected among themselves than with nodes from other communities \cite{barabasi2013network,fortunato2010}. For the identification of non-overlapping communities, we employ the modularity optimization approach \cite{newman2004}, chosen for its foundation in defining a null model that serves as a reference to evaluate the network's structure. Modularity acts as a metric to assess the quality of the identified partition. Indeed, partitions with high modularity have, respectively, dense/sparse connections of nodes within/between their clusters. \\
Consider a network with $N$ nodes and the associated adjacency matrix $A\in \mathbb{R}^{N\times N}$. In the context of an undirected unweighted graph, the entries of the adjacency matrix, $a_{ij}$, are such that $a_{ij}=1$ if a link exists between nodes $i$ and $j$, and $0$ otherwise. Our goal is to find non-overlapping communities represented by an $N$-dimensional vector $\eta$, where the $i$-th component $\eta_i$ indicates the set to which node $i$ belongs, as defined in \cite{macmahon2015community}.
The modularity, denoted as $Q(\eta)$, is defined as follows:
\begin{equation}
    Q(\eta) = \frac{1}{A_{tot}}\sum_{i,j}\left[A_{ij} - \langle A_{ij} \rangle \right]\delta(\eta_i,\eta_j)\,, \label{eq:modularity}
\end{equation}
where $\delta(\eta_i, \eta_j)$ is the Kronecker delta function equal to $1$ if $\eta_i=\eta_j$, and $0$ otherwise, signifying that only nodes within the same community are considered. $A_{tot}=\sum_{i,j}A_{ij}=2l$ is twice the total number of links $l$, and $\langle A_{ij} \rangle$ represents the employed null model. Traditionally, it is the so-called configuration model, in which $\langle A_{ij} \rangle = \frac{k_ik_j}{2l}$, with $k_i$ as the degree of node $i$ \cite{fortunato2010}. \\
In the presence of finite time series data having a global mode in the correlation matrix, the modularity can be expressed as follows:
\begin{eqnarray}
    Q(\eta) &=& \frac{1}{C_{norm}}\sum_{i,j}\left[C_{ij} - C^{(r)}_{ij}-C^{(m)}_{ij} \right]\delta(\eta_i,\eta_j) = \frac{1}{C_{norm}}\sum_{i,j} C^{(g)}_{ij}\delta(\eta_i,\eta_j)\,, \label{eq:modularity_RMT}
\end{eqnarray} 
where $C^{(r)}$, $C^{(m)}$, and $C^{(g)}$ represent the noise, market, and significant correlation components, respectively, as defined in equations (\ref{eq:noise_component}), (\ref{eq:market_component}), and (\ref{eq:real_component}). Additionally, $C_{norm}=\sum_{i,j}C_{ij}$. Research such as \cite{utsugi2004rmt,potters2005rmt,plerou2002rmt} has demonstrated that the eigenvector components of $C^{(g)}$ exhibit alternating signs, allowing for the identification of groups of stocks influenced by similar factors. This provides an effective basis for community detection in financial networks.\\
The modularity $Q(\eta)$ assumes values within the interval $[-0.5,1]$, indicating the edge density within communities relative to edges between communities. Higher modularity values denote a stronger community structure, with nodes forming distinct clusters, while lower values suggest a more uniform distribution of edges across the network.

\subsection{Signature}
The concept of signature derives from the field of path theory, providing a structured and comprehensive representation of the temporal evolution within a time series. Its potency lies in capturing both temporal and geometric patterns embedded within the time series. Temporal patterns encompass long-term dependencies and recurrent trends across time, while geometric patterns encompass the shape of time series trajectories, and intricate data behaviors such as loops and self-intersections \cite{lyons2014path}. \\
For the sake of clarity, we shall adopt the notation presented in \cite{ni2020conditional} and restrict our discussion to continuous functions mapping from a compact time interval $J:=[a,b]$ to $\mathbb{R}^d$ with finite $p$-variation, all starting from the origin. This space is denoted as $C^{p}_0(J, \mathbb{R}^d)$.\\ Let $\mathrm{T}((\mathbb{R}^{d})):=\oplus_{k=0}^{\infty}(\mathbb{R}^{d})^{\otimes k}$ represent a tensor algebra space, encompassing the signatures of $\mathbb{R}^{d}$-valued paths, offering their comprehensive representation. Furthermore, let $S_i = \{s_i(1), s_i(2), \dots, s_i(T)\}$ denote a discrete time series. To bridge the gap between the discrete and continuous cases, the time series needs to be converted into a continuous path, achieved through methods like the lead-lag transformation or the time-join transformation \cite{levin2013learning}. Let $L$ be the continuous path produced by the lead-lag transformation, which we adopt due to its ability to directly extract various features including path volatility (which comes from the second term of the signature), a crucial facet in finance. Consequently, we define the signature $\mathcal{S}$ and the truncated signature at level $M$, denoted as $\mathcal{S}_M$, as follows:
\begin{definition}[Signature and Truncated Signature]
Let \(L \in C^{p}_0(J, \mathbb{R}^d)\) be a path. The signature \(\mathcal{S}\) of the path \(L\) is defined as:
\begin{equation}
    \mathcal{S} = (1, L_{J}^{1}, \dots, L_{J}^{k}, \dots ) \in \mathrm{T}((\mathbb{R}^{d})\,, \label{eq:signature}
\end{equation} 
where \(L_{J}^{k}=\int_{t_1\ < t_2 < \dots t_k, t_1, \dots t_k \in J}{dL_{t_{1}}\otimes \dots \otimes dL_{t_{k}}}\) are called iterated integrals. \\
The truncated signature of degree \(M\) is defined as:
\begin{equation}
    \mathcal{S}_M = (1, L_{J}^{1}, \dots, L_{J}^{M} )\,. \label{eq
    :truncated_signature}
\end{equation} 
\end{definition}
The signature structure offers a hierarchical interpretation, with lower-order components capturing broad path attributes and higher-order terms  revealing intricate characteristics (including higher-order moments, and local geometric features). Importantly, the signature remains invariant under reparameterization, preserving integral values despite time transformations. It also adheres to translation invariance and concatenation properties \cite{chen1958integration}. The truncated signature preserves the first $\frac{d^{M+1}-1}{d-1}$ iterated integrals, with $M$ denoting truncation degree and $d$ representing path dimension. The factorial decay of neglected iterated integrals ensures minimal information loss in truncation of $\mathcal{S}$ \cite{lemercier2021distribution}.\\
Given two stochastic processes, $A$ and $B$, defined on 
a probability space $(\Omega, \mathbb{P}, \mathcal{F})$, and supposing equation (\ref{eq:signature}) holds almost surely for both $A$ and $B$, with expected values of $\mathcal{S}(A)$ and $\mathcal{S}(B)$ being finite, we have the following theorem \cite{lyons2015expected}: \\
\begin{theorem} [Expected Signature] \label{th:expected_sig}
Let $A$ and $B$ be two $C_{0}^{1}(J,\mathbb{R}^{d})$-valued random variables. If $\mathbb{E}[\mathcal{S}(A)]=\mathbb{E}[\mathcal{S}(B)]$, and $\mathbb{E}[\mathcal{S}(A)]$ has infinite radius of convergence, then $A \overset{d}{=} B$, i.e., $A$ and $B$ are equal in distribution.  
\end{theorem}
The signature uniquely defines a path's trajectory \cite{lyons1998}, under suitable assumption, while the expected signatures uniquely determine the distributions of paths, paralleling the role of moment generating functions \cite{chevyrev2016characteristic}. For a more comprehensive elaboration, rigorous formulations, and visual examples, consult \cite{lyons2014path,levin2013learning,chevyrev2016primer}.

\section{Signature-based Similarity Matrix} 
\label{section_signature}
In our research, we introduce a novel approach that replaces the conventional correlation matrix, $C$, with a similarity matrix derived from the truncated signature of the logarithmic increments of each time series 
$S_i$. This novel concept is rooted in the uniqueness of the signature, which can be likened to the moment generating function, making it an ideal candidate for quantifying similarity between time series. The hypothesis here is that if two time series possess highly similar signatures, they should exhibit substantial similarity in their behaviors.\\
To construct this similarity matrix, we embark on a multi-step process. First, we apply the lead-lag transformation to the logarithmic increments of each time series $S_i$, yielding the path $L_i$. Subsequently, we compute the truncated signature, denoted as $\mathcal{S}_M$, with a truncation degree $M$ set to $3$, applied to $L_i$
, also denoted by $\mathcal{S}_M(L_i)$. We then proceed to compute a similarity measure between each pair of stocks based on their truncated signatures. Three distinct 
measures are employed for this purpose: Euclidean Distance (ED), Cosine Similarity (CS), and Radial Basis Function (RBF) kernel.
The choice of these measures is deliberate: ED is selected for its sensitivity to data magnitude and computational efficiency, CS for its scale invariance, resistance to outliers, and suitability for time series trend analysis, and the RBF kernel for its capability to capture complex non-linear relationships. Moreover, we convert the Euclidean distance (which is a dissimilarity metric) into a similarity metric using a strictly monotone decreasing function, specifically $f(x)=\frac{1}{a+x}$ with $a=1$.\\
Finally, we obtain a similarity matrix $P$, with entries $p_{ij} \in [0,1]$, where a value of 1 signifies that elements $i$ and $j$ are perfectly alike. Following the creation of this similarity matrix, we subject it to filtering processes, specifically the threshold method and RMT, as elaborated upon in Section \ref{section_3}. This filtering serves the purpose of retaining only the significant similarities among the time series, thereby enhancing the robustness and effectiveness of our approach.

\section{Experimental Evaluation}
\label{section_4}

\begin{figure}[t]
\centering
\includegraphics[width=0.75\textwidth]{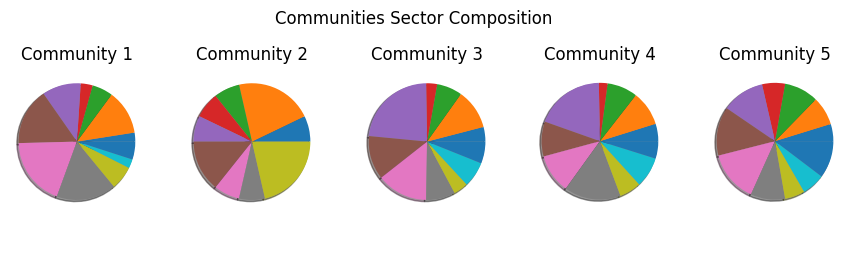}
\caption{Communities identified using the Louvain Community Detection Algorithm on the signature-based Similarity Matrix obtained via the Euclidean Distance and subsequently filtered using RMT. Communication Services ({\fboxsep=0pt\fbox{\color{MidnightBlue}\rule{3.5mm}{2mm}}}), Consumer Discretionary ({\fboxsep=0pt\fbox{\color{Orange}\rule{3.5mm}{2mm}}}), Consumer Staples ({\fboxsep=0pt\fbox{\color{ForestGreen}\rule{3.5mm}{2mm}}}), Energy ({\fboxsep=0pt\fbox{\color{Red}\rule{3.5mm}{2mm}}}), Financials ({\fboxsep=0pt\fbox{\color{Fuchsia}\rule{3.5mm}{2mm}}}), Health Care ({\fboxsep=0pt\fbox{\color{Brown}\rule{3.5mm}{2mm}}}), Industrials ({\fboxsep=0pt\fbox{\color{Rhodamine}\rule{3.5mm}{2mm}}}), Information Technology ({\fboxsep=0pt\fbox{\color{Gray}\rule{3.5mm}{2mm}}}), Materials ({\fboxsep=0pt\fbox{\color{GreenYellow}\rule{3.5mm}{2mm}}}) , Real Estate ({\fboxsep=0pt\fbox{\color{Aquamarine}\rule{3.5mm}{2mm}}}), and Utilities ({\fboxsep=0pt\fbox{\color{NavyBlue}\rule{3.5mm}{2mm}}}).}
\label{fig:community}
\end{figure}

\begin{figure}[t]
\centering
\includegraphics[width=0.67\textwidth]{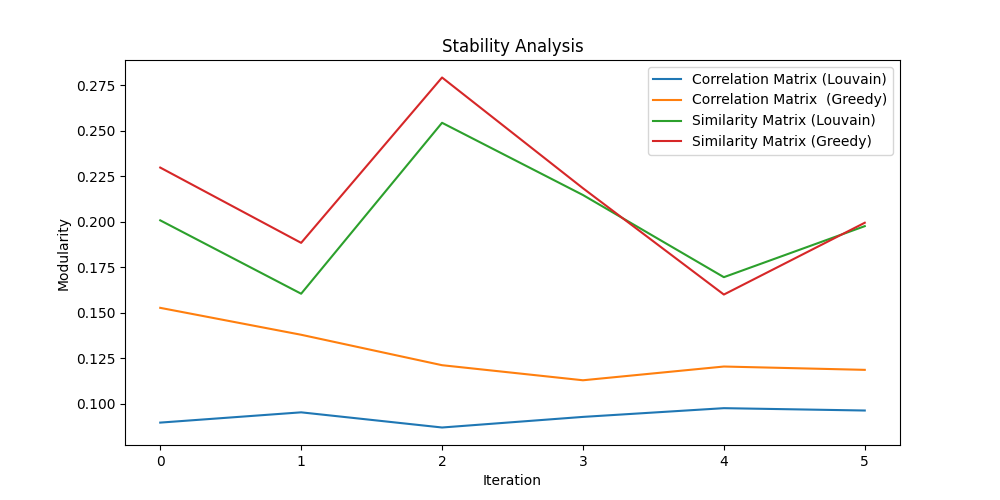}
\caption{Stability analysis for the correlation matrix and the signature-based similarity matrix for the Standard and Poor's 500. }
\label{fig:stability}
\end{figure}

The principal aim of this study is to showcase a substantial enhancement in the modularity metric when replacing the traditional correlation matrix with a similarity matrix derived from signatures in the context of community detection. A higher modularity score signifies an improved capability of the algorithm to identify more cohesive and distinguishable communities in the dataset. Additionally, our investigation highlights that the identified communities do not rigidly align with the initially assigned data categories.\\
To conduct this analysis, we focus our attention to the S\&P500 stock exchange, a market encompassing 500 major publicly traded companies spanning diverse sectors and industries in the United States market \cite{sp500}. Notably, this index classifies each stock into one of eleven distinct sectors: Communication Services, Consumer Discretionary, Consumer Staples, Energy, Financials, Health Care, Industrials, Information Technology, Materials, Real Estate, and Utilities.
Our data collection process starts from \newdate{date_start_out}{10}{07}{2010} \displaydate{date_start_out} to \newdate{date_end_out}{10}{07}{2023} \displaydate{date_end_out}. After computing the logarithmic returns and eliminating stocks with insufficient data, our dataset encompasses $443$ stocks for analysis, each comprising $3720$ observations. Consequently, we denote $N=443$ as the number of stocks and $T=3720$ as the length of each time series.\\
In this study, we use as baseline models the correlation matrix filter with a predefined threshold and the RMT-based filter, as detailed in Section \ref{section_3}. Within the threshold method, we determine the optimal threshold value following the procedure in \cite{macmahon2015community}. Subsequently, we retain correlation entries exceeding the threshold of $0.0437$.
The eigenvalue distribution under the RMT framework confirms the presence of the market model. Specifically, the largest eigenvalue of the correlation matrix is approximately $174$, with the second-largest eigenvalue around $20$. Consequently, we apply the filtering process to the correlation matrix as defined in equation (\ref{eq:real_component}). Notably, the market mode is also observed in the similarity matrix based on signatures.\\
Table \ref{tab:res1} presents the modularity results for the analyzed models, utilizing both the Louvain Community Detection Algorithm and the Clauset-Newman-Moore Greedy Modularity Maximization Algorithm.
\begin{table}[t]
\centering
\resizebox{0.67\textwidth}{!}{\begin{tabular}{|l|l|l|l|l|l|}
\hline
Data Type                                                                                    & Filtering Method & \begin{tabular}[c]{@{}l@{}}Modularity \\ (Louvain)\end{tabular} & \begin{tabular}[c]{@{}l@{}}Modularity\\  (Greedy)\end{tabular} & \begin{tabular}[c]{@{}l@{}}\# Cluster \\ (Louvain)\end{tabular} & \begin{tabular}[c]{@{}l@{}}\# Cluster \\ (Greedy)\end{tabular} \\ \hline
\multirow{2}{*}{\begin{tabular}[c]{@{}l@{}}Correlation\\ Matrix\end{tabular}}                & Threshold        & $0.0207$ & $0.0159$  & $81$    & $11$   \\ \cline{2-6}   & RMT  & $0.0987$    & $0.1185$  &  $4$  & $11$ \\ \hline
\multirow{6}{*}{\begin{tabular}[c]{@{}l@{}}Signature-based\\ Similarity Matrix\end{tabular}} & Threshold\_ED    & $0.1796$     & $0.1783$  & $8$   & $9$  \\ \cline{2-6} 
    & Threshold\_CS    & $0.0020$   & $-0.0134$    & $430$  & $11$    \\ \cline{2-6} 
    & Threshold\_RBF   & $0.0976 $   & $0.0952$    & $112$  & $11$    \\ \cline{2-6} 
    & RMT\_ED          & $0.1975$ & $0.1994$  & $5$   & $11$     \\ \cline{2-6} 
     & RMT\_CS          & $0.1177$    & $0.8527$      & $2$      & $11$  \\ \cline{2-6} 
    & RMT\_RBF         & $0.1326$     & $0.1428$        & $4$     & $11$        \\ \hline
\end{tabular}}
\caption{Modularity results for the correlation matrix and for the signature-based similarity matrix for the Standard and Poor's 500.}
\label{tab:res1}
\end{table}
Utilizing a signature-derived similarity matrix consistently yields higher modularity values, indicating superior performance in identifying more distinct communities than the correlation matrix approach. The exception is when applying threshold-based filtering to the cosine similarity matrix. Furthermore, results are more consistent when using similarity matrices based on the Euclidean distance, whereas outcomes from the cosine similarity matrix vary significantly in terms of modularity values and cluster count.\\
Remarkably, the Louvain algorithm, when applied to the threshold-filtered matrix, identifies numerous smaller communities, each comprising a single stock. This phenomenon holds for both the correlation matrix and the signature-based similarity matrix, except for the Euclidean distance case.\\
Figure \ref{fig:community} illustrates the community structure generated by the Euclidean distance-based similarity matrix, filtered using RMT. The algorithm identifies a total of $5$ communities, in contrast to the S\&P$500$ index that classifies these stocks into $11$ distinct categories. All $5$ algorithm-identified communities encompass stocks from different sectors according to the S\&P$500$ classification, suggesting concealed correlations among stocks from various sectors, highlighting intricate inter-sector relationships.\\
To evaluate the robustness of our proposed methodology, we conducted a stability analysis. This entailed gradually increasing the number of observations in the dataset, starting with roughly one-third of the original observations for the 443 stocks. We incrementally added observations until reaching the dimensions of the original dataset. This analysis aimed to demonstrate that our method's effectiveness remains consistent regardless of the quantity of observations considered. It is important to note that RMT requires only that the number of observations ($T$) exceeds the number of stocks ($N$). The results of this stability analysis are presented in Figure \ref{fig:stability}.
In this study, we primarily examine the similarity matrix derived from the Euclidean distance via the signature, given its previously demonstrated superior performance in achieving higher modularity. We also include the conventional correlation matrix in our analysis, subjecting both matrices to RMT-based filtration. Interestingly, even within this context, the modularity consistently exhibits higher values when utilizing the signature-based similarity matrix for community detection. Notably, the performance of the greedy algorithm for community detection within the signature-based approach appears to be influenced by the volume of observations.

\section{Conclusion}
\label{section_5}
This study explores contemporary techniques for filtering correlation matrices in community detection. We introduce a novel approach, substituting the correlation matrix with a signature-derived similarity matrix. We evaluate three similarity measures: nonlinearly transformed Euclidean distance, cosine similarity, and Radial Basis Function (RBF) similarity. Using the S\&P$500$ dataset, we empirically assess this method's performance with the modularity metric. Results consistently indicate enhanced modularity, with the Euclidean distance-based similarity matrix performing the best.\\
Future research will involve in-depth exploration of the structural properties in community detection using the signature-based similarity matrix. In particular, we plan to: investigate the reasons behind the higher modularity achieved by the proposed method with respect to other similarity measures; extend it by directly computing the signature of the vector of (paths derived from) time series; explore its integration into portfolio optimization and risk management strategies; apply it to other contexts involving time series, such as movement analysis.

\section*{Acknowledgment}
The authors were partially supported by the PRIN 2022 project ``Multiscale Analysis of Human and Artificial Trajectories: Models and Applications'', funded by MUR (CUP: D53D23008790006).

%

\end{document}